\documentclass{IEEEtaes}

\usepackage[utf8]{inputenc}
\usepackage{dsfont}
\usepackage{color,array,amsthm,cite,multicol}
\usepackage{graphicx}
\usepackage{float}

\usepackage{amsmath,amssymb,bm,bbm,latexsym,subcaption,epstopdf}
\usepackage[version=4]{mhchem}
\usepackage[english]{babel}

\usepackage{hyperref}
\usepackage[dvipsnames]{xcolor}

\jvol{XX}
\jnum{XX}
\jmonth{XXXXX}
\paper{1234567}
\pubyear{2026}
\doiinfo{TAES.2026.Doi Number}

\graphicspath{{./figures/}} 
\DeclareGraphicsExtensions{.pdf,.eps}

\newcommand{\dfn}{\triangleq}
\newcommand{\given}{\mid}

\DeclareMathOperator{\sign}{sgn}
\DeclareMathOperator{\sat}{sat}

\newcommand{\meas}{\ensuremath{\mathcal{Y}}}

\DeclareMathOperator{\erf}{\text{erf}}

\let\abs=\envert

\newcommand{\lr}{R_{\text{SK}}}
\newcommand{\reff}{R_{\mathrm{eff}}}

\newtheorem{proposition}{Proposition}

\newtheorem{remark}{Remark}

\newcommand{\fullref}[2]{\ref{#1}.\ref{#2}}   

\newif\ifmarkup
\markupfalse 
\protected\long\def\rev#1{\ifmarkup{\color{blue}#1}\else{#1}\fi} 

\ifmarkup
  \usepackage[switch,pagewise]{lineno}
  
  \newcommand*\patchAmsMathEnvironmentForLineno[1]{%
    \expandafter\let\csname old#1\expandafter\endcsname\csname #1\endcsname
    \expandafter\let\csname oldend#1\expandafter\endcsname\csname end#1\endcsname
    \renewenvironment{#1}%
      {\linenomath\csname old#1\endcsname}%
      {\csname oldend#1\endcsname\endlinenomath}}%
  \newcommand*\patchBothAmsMathEnvironmentsForLineno[1]{%
    \patchAmsMathEnvironmentForLineno{#1}%
    \patchAmsMathEnvironmentForLineno{#1*}}%
  \AtBeginDocument{%
    \patchBothAmsMathEnvironmentsForLineno{equation}%
    \patchBothAmsMathEnvironmentsForLineno{align}%
    \patchBothAmsMathEnvironmentsForLineno{flalign}%
    \patchBothAmsMathEnvironmentsForLineno{alignat}%
    \patchBothAmsMathEnvironmentsForLineno{gather}%
    \patchBothAmsMathEnvironmentsForLineno{multline}}
  \linenumbers
\fi

\begin{document}

\title{Kill-Probability-Maximization Guidance: Breaking from the
  Miss-Distance-Minimization Paradigm}

\author{Liraz Mudrik}
\member{\rev{Member, IEEE}}

\author{Yaakov Oshman}
\member{Fellow, IEEE}
\affil{\rev{Technion--Israel Institute of Technology, Haifa 3200003, Israel}}

\receiveddate{Manuscript received XXXXX 00, 0000; revised XXXXX 00, 0000; accepted XXXXX 00, 0000.}

\corresp{{\itshape (Corresponding author: Liraz Mudrik)}. }

\authoraddress{\rev{The authors are with the Stephen B.\ Klein
Faculty of Aerospace Engineering, Technion--Israel Institute of
Technology, Haifa 3200003, Israel. Email: liraz109@gmail.com
(L.~Mudrik); yaakov.oshman@technion.ac.il (Y.~Oshman).}
This work is part of Dr.\ Mudrik's doctoral research, performed at the Technion.}


\markboth{MUDRIK AND OSHMAN}{KILL-PROBABILITY-MAXIMIZATION GUIDANCE}
\maketitle

\begin{abstract}
  Classical guidance laws aim at minimizing the miss distance, thus
  implicitly determining the minimum warhead lethality radius required
  against nominal targets.  However, nonnominal targets or scenarios
  might render the designed warhead insufficient, causing a
  significant degradation in the single-shot kill probability (SSKP).
  We propose a guidance methodology that shifts the interceptor's
  objective from minimizing the miss distance to directly maximizing
  the SSKP, while taking into account the warhead's probabilistic
  lethality model.  Complying with the generalized separation theorem,
  the paradigm presented here \rev{modifies} deterministic
  differential-game-based guidance laws using Bayesian decision
  theory.  \rev{For this purpose, we obtain the saddle-point solution
  of the deterministic differential game whose objective is the
  SSKP.}  Extensive Monte Carlo simulations demonstrate \rev{substantial}
  SSKP improvement over the standard and recently introduced
  estimation-aware guidance laws, when tested against nominal and
  nonnominal evasively maneuvering targets.
\end{abstract}

\newpage
\section{Introduction}
Classical guidance laws have been traditionally developed using
various optimization techniques, in particular optimal control and
differential game theories, typically assuming perfect information
about both the interceptor and the target
states~\cite{zarchan_tactical_2012,ben-asher_advances_1998}.  In
practice, however, the information pattern is never perfect, and, in
stochastic scenarios, hit-to-kill (HTK) capability cannot be
achieved~\cite{shaviv_estimation-guided_2017}, which necessitates the
use of a warhead.  For example, the differential game guidance law
(DGL1)~\cite{gutman_optimal_1979} guarantees HTK performance under
linearized, deterministic conditions~\cite{shinar_solution_1981}, yet
its performance severely degrades in stochastic
scenarios~\cite{shinar_what_2003}, and\cite{shaferman_stochastic_2016}
demonstrates the necessity of a warhead even when using such
state-of-the-art guidance laws.

Several key factors affect the design of the required warhead: the
type of the nominal target, e.g., an airborne vehicle or a ballistic
missile; the lethality model of the warhead; the guidance law used by
the pursuer against the nominal target; and the single shot kill
probability (SSKP) that the warhead and the guidance law should
provide against the nominal target.  Since the nominal target is
prespecified, and assuming that the SSKP requirement is given, the
design of the warhead reduces to properly choosing the optimal
combination of the warhead's lethality model and the interceptor's
guidance law that satisfies the SSKP requirement.  Commonly assumed to
be symmetric, the lethality model of a given warhead provides the
probability of killing the target as a function of the miss distance;
a thorough description of such models is presented
in~\cite{jaiswal_search_1997}.  The standard approach models the
warhead's lethality using the cookie-cutter (CC) damage
function~\cite{shinar_integrated_2007}, which, unrealistically,
assumes a hard threshold: miss distances below it yield a hit, whereas
those above it yield a miss. Thus, adopting the standard CC model
renders the miss distance the focus of the design process. Naturally,
then, the objective of the guidance law becomes to achieve, in some
statistical sense, a miss distance that would satisfy the SSKP
requirement. In practical terms, Monte Carlo (MC) simulations are run
against the nominal target, in which the miss distance outcome of each
simulation is a realization of the miss distance random
variable. Based on these simulations, the empirical miss distance
cumulative distribution function (CDF) is evaluated, which enables
choosing the CC damage function's threshold so as to satisfy the SSKP
requirement.

The design process outlined above is predicated on the concept of a
\emph{nominal target}, i.e., the particular target against which the
pursuer is calibrated to achieve the specified SSKP requirement.
However, the caveat is that the robustness of this approach cannot be
guaranteed.  When facing \emph{nonnominal} targets in practical
scenarios, the resulting warhead/guidance law combination may prove
inadequate in achieving the required SSKP. The closest prior work
addressing a related limitation is in the deterministic setting:
\cite{weiss_minimum_2016} replaces miss-distance minimization with a
hard miss-distance constraint, but assumes perfect information,
precluding its application in stochastic, nonnominal
scenarios. Responding to these limitations, in this paper we propose a
novel design approach that robustifies interception performance by
incorporating the warhead's lethality model into the guidance
strategy, and directly maximizing the SSKP against any target the
interceptor may encounter, whether nominal or not.

The new approach is based on two main pillars. First, we propose a
soft, probabilistic damage function that assigns a kill probability to
each miss distance, thereby capturing the gradual degradation of
interception performance as the miss distance increases and yielding a
lethality model which is more realistic than models employing hard
miss distance thresholds. A smooth lethality model has also been
employed in~\cite{ShinarForteKantor1994} in the context of mixed
strategy guidance, although without incorporating the model into the
real-time guidance command.  \rev{As shown in the sequel, incorporating the
probabilistic lethality model into the cost of the underlying
deterministic game leaves that game's saddle-point strategies
unaltered.  The lethality model therefore affects the guidance command
only under imperfect information, through the posterior PDF of the
game state.} Second, we employ the guidance law design
paradigm recently introduced in \cite{Mudrik2026unified} and
\cite{Mudrik2026comprehensive}, which modifies deterministic
differential game-based guidance laws\rev{, called the DGL laws,} using
Bayesian decision theory
to yield improved stochastic interception performance. Complying with
the generalized separation theorem
(GST)~\cite{striebel_sufficient_1965,witsenhausen_separation_1971},
this approach yields estimation-aware (EA) guidance laws that can cope
with imperfect information, nonlinear dynamics, and possibly
non-Gaussian distributions. It does so by employing the interacting
multiple model particle filter (IMMPF)~\cite{blom_exact_2007}, which
approximates the entire posterior probability density function (PDF),
as required by the Bayesian decision algorithm. Incorporating the
probabilistic lethality model into the Bayesian decision criterion,
and exploiting the game space decomposition provided by these guidance
laws, enables making optimal decisions over the posterior PDF, thus
maximizing directly the resulting SSKP.

The remainder of this paper is organized as follows.
Section~\ref{Prob_Form} formally states the problem.
Section~\ref{sec:lethality} discusses kill probability evaluation,
warhead lethality models, and warhead design approaches.  The guidance
strategy that incorporates the lethality model is derived in
Section~\ref{sec:SSKP_awa_GL}.  Section~\ref{sec:sim} presents the
results of an extensive Monte Carlo simulation study.  Concluding
remarks are offered in Section~\ref{sec:concl}.

\section{Problem Statement}
\label{Prob_Form}

\subsection{Nonlinear Kinematics and Dynamics}
\begin{figure}[tbh]
\centering
\includegraphics[width=0.45\textwidth]{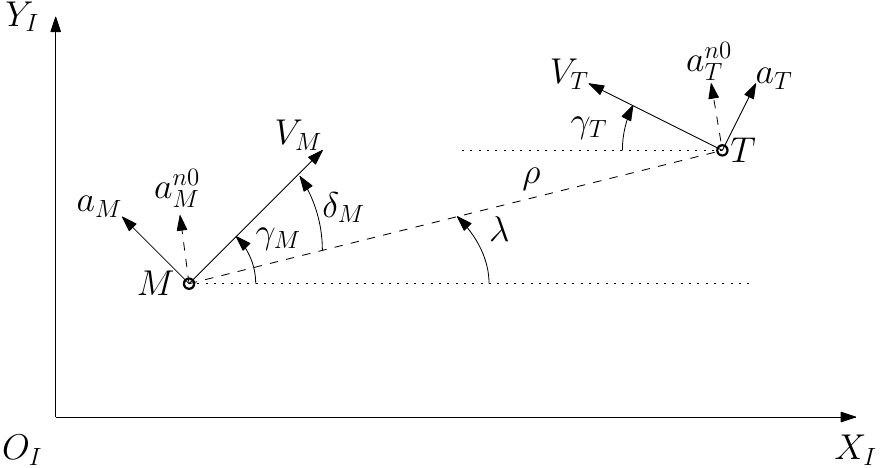}
\caption{Planar engagement geometry.}
\label{fig:Planar-engagement-geometry}
\end{figure}

A single-pursuer, single-evader interception scenario is considered. 
Figure~\ref{fig:Planar-engagement-geometry} shows a schematic view of the geometry of the assumed planar endgame scenario, where $X_{I}$-$O_{I}$-$Y_{I}$ is a Cartesian inertial reference frame. 
The interceptor and the target are denoted by $M$ and $T$, respectively, and variables associated with each are distinguished by corresponding subscripts. 
The speed, normal acceleration, and path angle are denoted by $V$, $a$, and $\gamma$, respectively.  
The slant range between the interceptor and the target is $\rho$, and the line-of-sight (LOS) angle, measured with respect to the $X_{I}$ axis, is $\lambda$.

The following assumptions are adopted: both players are modeled as point masses; the interceptor's path angle and lateral acceleration are known via its own navigation system; both speeds $V_{M}$ and $V_{T}$ are known and time-invariant; both players possess first-order dynamics with known time constants $\tau_{M}$ and $\tau_{T}$; the lateral acceleration bounds $a_{M}^{\max}$ and $a_{T}^{\max}$ are known constants; and the target selects its acceleration command $u_{T}$ from a known set of $R$ possible evasion maneuvers.
Following common practice, we use polar coordinates to define the equations of motion (EOM). The interceptor's state vector used for estimation is:
\begin{equation}
\textbf{x}_{M} = \begin{bmatrix}
  \rho & \lambda & \gamma_{T} & a_{T}
\end{bmatrix}^{T}
\label{eq:StateVec1}
\end{equation}
and the corresponding EOM are:
\begin{subequations}
\begin{align}
	\dot{\rho}       & =  V_{\rho}                                       \\
	\dot{\lambda}    & =  \frac{V_{\lambda}}{\rho}                       \\
	\dot{\gamma}_{T} & =  \frac{a_{T}}{V_{T}}                            \\
	\dot{a}_{T}      & =  -\frac{a_{T}}{\tau_{T}}+\frac{u_{T}}{\tau_{T}}
\end{align}
\label{eq:2}
\end{subequations}
where
\begin{subequations}
\begin{align}
	V_{\rho}    & =  - (V_{M}\cos\delta_{M}+V_{T}\cos\delta_{T}) \\
	V_{\lambda} & = -V_{M}\sin\delta_{M}+V_{T}\sin\delta_{T}     \\
	\delta_{M}  & = \gamma_{M}-\lambda, \quad \delta_{T} = \gamma_{T}+\lambda.
\end{align}
\label{eq:EOM1}
\end{subequations}

Based on the underlying assumptions, the interceptor's path angle and lateral acceleration satisfy the following evolution equations:
\begin{subequations}
	\label{eq:EOM2}
	\begin{align}
		\dot{\gamma}_{M} & =  \frac{a_{M}}{V_{M}}                            \\
		\dot{a}_{M}      & =  -\frac{a_{M}}{\tau_{M}}+\frac{u_{M}}{\tau_{M}}
	\end{align}
\end{subequations}
where $u_{M}$ is the interceptor's acceleration command.

\subsection{Measurement Model}
We assume that the interceptor can measure the bearing angle $\delta_{M}$ between its own velocity vector and the LOS to the target, yielding the following measurement equation
\begin{equation}
y = \delta_{M}+\nu = \gamma_{M}-\lambda+\nu\label{eq:Meas}
\end{equation}
where $\nu$ is the (possibly non-Gaussian) measurement noise.

\subsection{Linearized Model}
Differential game-based guidance laws are commonly derived using linear models~\cite{glizer_complete_2008}. Let $\xi$ denote the relative separation normal to the initial LOS. Assuming a near-collision course (small angles), we define the normalized time-to-go and the nondimensional state vector as
\begin{equation}
	\tau = t_{go} / \tau_{M}, \qquad \tau_{0} = t_{f} / \tau_{M},
\end{equation}
\begin{equation}
	\bar{\textbf{x}}(\tau) = 
	\begin{bmatrix}
		\frac{x_{1}(\tau)}{a_{T}^{\max} \tau_{M}^{2}} & 
		\frac{x_{2}(\tau)}{a_{T}^{\max} \tau_{M}} & 
		\frac{x_{3}(\tau)}{a_{M}^{\max}} & 
		\frac{x_{4}(\tau)}{a_{T}^{\max}}
	\end{bmatrix}^{T},
\end{equation}
where $t_{go} \dfn t_{f} - t \approx -\rho/V_{\rho}$ and $\textbf{x} = [\xi,\, \dot{\xi},\, a_{M},\, a_{T}]^{T}$. The normalized controls $\bar{u}$, $\bar{v}$ satisfy $|\bar{u}|, |\bar{v}| \leq 1$, and the nondimensional EOM are
\begin{subequations}
\label{eq:lin_sys}
	\begin{align}
		d\bar{x}_{1}/d\tau & = - \bar{x}_{2}, & \bar{x}_{1}(\tau_{0}) = 0 \;\;\; \\
		d\bar{x}_{2}/d\tau & = - \bar{x}_{4} + \mu \bar{x}_{3} , & \bar{x}_{2}(\tau_{0}) = \bar{x}_{20} \\
		d\bar{x}_{3}/d\tau & = \bar{x}_{3} - \bar{u},  & \bar{x}_{3}(\tau_{0}) = 0 \;\;\; \\
		d\bar{x}_{4}/d\tau & = (\bar{x}_{4} - \bar{v}) / \epsilon ,  & \bar{x}_{4}(\tau_{0}) = 0 \;\;\;
	\end{align}
\end{subequations}
where $\mu = a_{M}^{\max} / a_{T}^{\max}$ and $\epsilon = \tau_{T} / \tau_{M}$ are the maneuverability and agility ratios, respectively, and 
\begin{equation}
    \bar{x}_{20} = (V_{T} \sin \gamma_{T_{0}} - V_{M} \sin \gamma_{M_{0}}) / (a_{T}^{\max} \tau_{M}).    
\end{equation}
In compact form,
\begin{equation}
\dot{\bar{\textbf{x}}}(\tau) = \textbf{A} \bar{\textbf{x}}(\tau) + \textbf{B}\bar{u}(\tau) + \textbf{C}\bar{v}(\tau).
\label{eq:compact_lin}
\end{equation}

\section{Warhead Lethality and Design}
\label{sec:lethality}
\subsection{Kill Probability Evaluation}
We measure the lethality of a given warhead by its probability of
killing the target.  Recall that both players are modeled as point
masses.  Hence, the probability of destroying a point target by
initiating a warhead at the final position of the interceptor over the
$X_{I}$-$O_{I}$-$Y_{I}$ plane is given by
\begin{equation}
	P_{d} = \int_{-\infty}^{\infty} \int_{-\infty}^{\infty} p(\bar{x},\bar{y}) d(\bar{x},\bar{y}) d\bar{x} d\bar{y},
	\label{eq:Pd}
\end{equation}
where $\bar{x} \dfn x-x^{f}_{M}$ and $\bar{y} \dfn y-y^{f}_{M}$ are
the relative coordinates with respect to the initiating point of the
warhead.  Here $d(\cdot,\cdot):\mathbb{R}^{2}\rightarrow[0,1]$ is the
damage probability function, which measures the probability of
damaging the target, and
$p(\cdot,\cdot): \mathbb{R}^{2}\rightarrow [0,\infty]$ is the PDF of
the target relative position over the plane.
A thorough review of damage functions appears
in~\cite{goethals_broadening_2015}.  For simplicity, we assume that
the damage function is a circularly symmetric function, that is, a
function of the radial distance only, where the radial distance is the
miss distance $M_{s} = \sqrt{\bar{x}^{2} + \bar{y}^{2}}$. We note that
the damage function is a random variable, as it is a function of the
random miss distance.

Calculating the double integral in~\eqref{eq:Pd} is, generally,
complicated, and analytical solutions are only possible in particular
cases.  One such case is when the PDF of the target relative position
at $t_{k}$, $(\bar{x}_{k},\bar{y}_{k})$, is modeled using the delta
function:
\begin{equation}
  \label{eq:p_delta}
  p(\bar{x},\bar{y}) \sim \delta (\bar{x}-\bar{x}_{k} , \bar{y}-\bar{y}_{k} ),
\end{equation}
rendering $P_{d}$ solely based on the damage function:
\begin{equation}
  \label{eq:Pd_delta}
	{P_d}_k = d(\bar{x}_{k},\bar{y}_{k}).
\end{equation}
      
Under an imperfect information pattern, the PDF of the target relative
position can only be estimated from measurements.  Whereas the Kalman
filter assumes linear dynamics and Gaussian noise distributions (thus
providing only first and second moment estimates), particle filters
(PFs) approximate the entire posterior PDF without such restrictive
assumptions.  Furthermore, because PFs represent the posterior using
weighted delta functions, they are naturally suited to
evaluating~\eqref{eq:Pd_delta} for each particle.  To best adapt to a
target capable of multiple evasion maneuvers, possibly following a
non-Markovian switching process, we adopt the
IMMPF~\cite{blom_exact_2007}.

The IMMPF runs a bank of PFs matched to all possible modes.  At each
time $t_{k}$ the PDF is represented by a set of particles and
associated weights
$\{ \textbf{x}_{k}^{r,s}, w_{k}^{r,s} \mid r\in\mathcal{R}, s\in \{1,
\dots, S\}\} $, where $\textbf{x}_{k}^{r,s}$ is the state vector of
particle $s$ and mode $r$, and $w_{k}^{r,s}$ is its associated weight.
$\mathcal{R}$ is the set of $R$ discrete modes, and $S$ is the number
of particles of each mode, rendering the total number of particles
$N_{p}=RS$.  The posterior PDF is approximated by
\begin{equation}
	p(\textbf{x}_{k} \given \meas_{k} ) \approx \sum_{r=1}^{R} \sum_{s=1}^{S} w_{k}^{r,s} \delta (\textbf{x}-\textbf{x}_{k}^{r,s} )
	\label{eq:postPDF}
\end{equation}
where $\meas_{k}$ is the history of measurements at time $t_{k}$.
Thus, based on \eqref{eq:p_delta} and \eqref{eq:Pd_delta}, for each
particle $j$ at time $t_{k}$, $P_{d}$ is readily evaluated using the
damage function:
\begin{equation}
  \label{eq:Pd_damage}
	{P_d}_k^j = d(\bar{x}_{k}^{j},\bar{y}_{k}^{j}).
      \end{equation}

\subsection{Lethality Models}
\subsubsection{Cookie-Cutter Lethality Model}
The CC damage function is probably the most used function, because of
its simple structure.  Essentially being the indicator random
variable, this function is given as
\begin{equation}
  \label{eq:cookie_damage}
	d(M_{s}) = \mathbbm{1}_{M_{s} \leq \lr} =
	\begin{cases}
		1, & M_{s} \leq \lr \\
		0, & M_{s} > \lr
	\end{cases},
\end{equation}
where $\lr$, the lethality radius, is the blast radius that ensures
a kill by the warhead.  Other typical options for modeling the damage
function are Gaussian, exponential, and Gamma damage
functions~\cite{jaiswal_search_1997,goethals_broadening_2015}.
\subsubsection{Probabilistic Lethality Model}      
Because the new guidance law, derived in the sequel, incorporates the
lethality model at its core, we next introduce a probabilistic
lethality model (PLM) that is more elaborate and realistic than the
commonly used, simplistic CC model.  Let $P_{m} \dfn 1 - P_{d}$ be the
miss probability.  The function
$P_{m}: \mathbb{R}_{+}\rightarrow [0,1]$ assigns a miss probability to
each miss distance, and is assumed to be known. Now, similarly to
option (d) of the alternative damage function models
in~\cite{goethals_broadening_2015}, let the damage function be modeled
as
\begin{equation}
  d(M_{s}) = \frac{1}{2} \left[ 1 - \erf \left( \frac{M_{s} - \mu_w}{\sqrt{2}\sigma_w} \right)  \right]
  \label{eq:d}
\end{equation}
where $\erf(\cdot)$ is the error function, and $\mu_w$ and $\sigma_w$ are
given warhead parameters. Using \eqref{eq:Pd_damage}, the miss
probability of each particle $j$ is, thus,
\begin{equation}
  P_{m}(M_{s}^{j}) = 1 - d(M_{s}^{j}) = \frac{1}{2} \left[ 1 + \erf \left( \frac{M^{j}_{s} - \mu_w}{\sqrt{2}\sigma_w} \right) \right]
	\label{eq:P_m}
\end{equation}
where $M_{s}^{j}$ is the miss distance estimate of particle $j$.  The
miss distance can be obtained for each particle using known solutions
for deterministic guidance laws, as shown in
Sec.~\fullref{sec:SSKP_awa_GL}{subsec:game_space} for the \rev{DGLs}.
Generalizing this result to other damage functions, e.g., those based
on exponential or Gamma distributions, is straightforward.

Because the PLM has no hard threshold, we define the effective lethal
radius of a warhead having the parameters $\mu_w$ and $\sigma_w$ as
\begin{equation}
  \label{eq:R_eff}
  \reff(n_\sigma) \dfn \mu_w - n_\sigma \sigma_w,
\end{equation}
where $n_\sigma > 0$ is a design parameter that sets the
kill-probability threshold associated with $\reff(n_\sigma)$. At
$M_s = \reff(n_\sigma)$, the corresponding kill probability is
$P_d[\reff(n_\sigma)] = \frac{1}{2}[1 + \erf(n_\sigma/\sqrt{2})] =
\Phi(n_\sigma)$, where $\Phi$ is the CDF of the standard normal
distribution; for example, $P_d[\reff(1)] \approx 0.8413$, whereas
$P_d[\reff(3)] \approx 0.9987$. Larger $n_\sigma$ tightens the
guaranteed-kill region, at the cost of a smaller $\reff$ for fixed
warhead parameters.

\begin{figure}[tbh]
  \centering
  \includegraphics[width=0.45\textwidth]{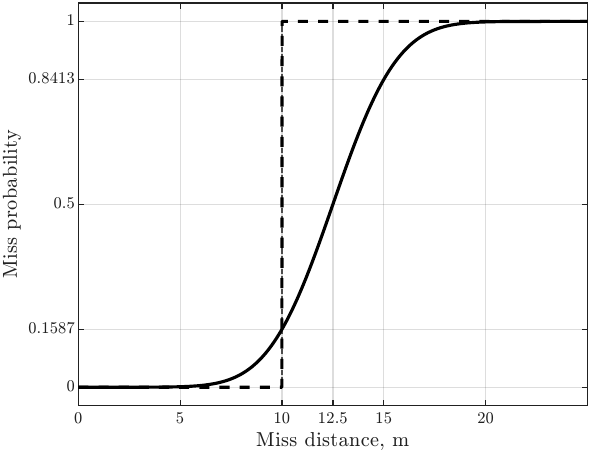}
  \caption{CC model (dashed; $\lr = 10$~m) vs PLM (solid;
    $\mu_w = 12.5$~m, $\sigma_w = 2.5$~m). The dash-dotted vertical
    line marks $\reff(1) = \lr$. The non-standard y-axis values denote
    $P_m$ evaluated at $\mu_w - \sigma_w$, $\mu_w$, and
    $\mu_w + \sigma_w$.}
  \label{fig:lethal_model}
\end{figure}

Figure~\ref{fig:lethal_model} compares the new PLM \eqref{eq:P_m} with
a CC model having a lethality radius of $\lr = 10$~m. The PLM's
parameters, $\mu_w = 12.5$~m and $\sigma_w = 2.5$~m, are chosen so
that the effective lethal radius $\reff(1) \dfn \mu_w - \sigma_w$
coincides with $\lr$, enabling a direct comparison of the two models
at the same design radius. The dash-dotted vertical line marks this
common radius, where the PLM attains a kill probability of
$P_d[\reff(1)] \approx 0.8413$. The y-axis is annotated at
$P_m(\mu_w - \sigma_w) \approx 0.1587$, $P_m(\mu_w) = 0.5$, and
$P_m(\mu_w + \sigma_w) \approx 0.8413$, which, together with the
x-axis tick at $\mu_w = 12.5$~m, expose the PLM's parametric
structure. As is clear from Fig.~\ref{fig:lethal_model}, whereas the
simplistic CC model assigns an unrealistic binary outcome to any
engagement (depending on the miss distance), the PLM realistically
expresses the gradual and monotonic degradation of warhead
effectiveness with increasing miss distance. This distinction between
the models is central to the proposed approach. By incorporating the
realistic PLM's profile into the guidance cost, the interceptor can
make decisions that account for actual miss probabilities, rather than
treating interception as a binary outcome process.
\subsection{Warhead Design}
\label{sec:performance}
The common approach to the design of guidance systems assumes that the
warhead possesses a CC damage function with a
to-be-determined lethality radius. Assuming a nominal target and an
appropriately chosen guidance law, the lethality radius is then
determined according to the statistical miss distance performance of
the guidance law, such that a prespecified SSKP criterion is met. We
first review this approach, before presenting the new guidance design
paradigm.
\subsubsection{Common Design Approach}
\label{sec:common_approach}
The common approach to determine the required warhead lethality radius
for a given guidance law and for a prespecified SSKP is described
in~\cite{shinar_integrated_2007}, where the CC damage function is
used.  A core component of this approach, the guidance law is
determined first, by solving a differential game having the miss
distance as its cost:
\begin{equation}
  J_{\text{MD}} = M_s. 
\end{equation}
Thus, the interceptor aims at driving the cost to zero and the target
aims at maximizing it. For the interceptor, the rationale for
minimizing the miss distance is clear, as a smaller miss would require
a lighter warhead in order to achieve a specified SSKP. Nevertheless,
we note that, importantly, both players completely disregard the
interceptor's warhead when deciding on their optimal strategies, as
the warhead's lethality radius would only be determined based on the
solution of the game. This renders the interceptor's guidance law
overly aggressive, as it aims at achieving a goal ($J_{\text{MD}}=0$)
which is not absolutely required when one considers 1) the existence
of a warhead, and 2) the realistic requirement of an SSKP strictly
smaller than one.
%
%

The procedure is then based on running an extensive MC simulation
study using this guidance law against nominal targets in various
nominal scenarios. This yields an empirical estimate of the miss
distance CDF, $\hat{F}_{M_{s}}(m)$, which satisfies
\begin{equation}
  \label{eq:CDF}
  \hat{F}_{M_s}(m) = \Pr(M_s \le m).
\end{equation}
Denote by $\mathcal{K}$ the event that the target is killed. Using the
law of total probability,
\begin{align}
  \label{eq:total}
  \Pr(\mathcal{K}) & = \Pr(\mathcal{K}\mid M_s > \lr)\Pr(M_s > \lr)\notag\\
                   & + \Pr(\mathcal{K}\mid M_s \le \lr)\Pr(M_s \le \lr),
\end{align}
and \eqref{eq:cookie_damage} readily yields
\begin{equation}
  \label{eq:prob_kill}
  \Pr(\mathcal{K}) = \Pr(M_s \le \lr) = \hat{F}_{M_s}(\lr).
\end{equation}
Setting $\Pr(\mathcal{K}) = \kappa$, the prespecified value of the
SSKP, then yields the required lethality radius of the warhead as the
solution of
\begin{equation}
  \label{eq:leth_radius}
  \hat F_{M_s}(\lr) = \kappa.
\end{equation}
%
%
\subsubsection{The New Approach}
Suppose now that the lethality model of the warhead has already been
determined via a common design procedure against some nominal target,
as outlined above. As the miss-distance-minimizing guidance law has
already been determined as well, when faced with a challenging,
nonnominal target, the combination of guidance law and its
corresponding warhead might fail to achieve the required SSKP
performance. This situation, then, calls for a new design approach, in
which the warhead's lethality model is incorporated into the guidance
strategy, and the new guidance law is devised with the objective of
directly maximizing the SSKP. Obviously, this renders the minimization
of miss distance immaterial, and the cost, which is to be minimized by
the guidance system, is set to be the miss probability, i.e.,
\begin{equation}
	\label{eq:J_Pm}
	J_\text{PLM} = P_{m}(M_{s}).
\end{equation}
%

\section{Kill-Probability-Maximization Guidance}
\label{sec:SSKP_awa_GL}
Complying with the GST conditions, we derive a new guidance law that
maximizes the interceptor's SSKP when its warhead lethality model is
given.  The derivation builds upon 1) the game space of deterministic
differential-game-based guidance laws, and 2) the posterior PDF
provided by the IMMPF algorithm.  For completeness, we first briefly
review the Bayesian decision theory framework. Next, we show how doing
away with the miss distance cost and focusing instead on the
miss-probability cost yields a kill-probability-maximization (KPM)
variant of any DGL-based law.  An illustrative example concludes the
section.
\subsection{Bayesian Decision Theory}
\label{subsec:bayes}
Following the Bayesian decision framework of~\cite{Mudrik2026unified,trees_detection_2004}, the optimal decision among $m$ hypotheses is obtained by minimizing the unnormalized additional risk $I_i$ (see~\cite{Mudrik2026unified} for its derivation from the conditional Bayesian risk), defined as
\begin{equation}
	I_{i} (\meas) \dfn \sum_{\substack{j=1 \\ j\neq i}}^{m} 
	P_{j} \, P(\meas \given H_{j}) \, (C_{ij} - C_{jj}),
	\label{Eq:I}
\end{equation}
where $C_{ij}$ is the cost of declaring that $H_{i}$ is true when
$H_{j}$ is true, $P_{j} = \Pr(H_{j}\text{ is true})$ is the prior
probability of hypothesis $H_j$, and $P(\meas \given H_j)$ is the
likelihood of the measurements given that $H_j$ is true. The optimal
decision rule is:
\begin{equation}
	\text{Decide } H_{i^\star}, \text{ where } i^{\star}
	\dfn \arg\min_{i\in \{1,\ldots,m\} }\{ I_{i}(\meas)\}.
	\label{Eq:Decision}
\end{equation}
\subsection{The DGL Game Space}
\label{subsec:game_space}
Our new approach is based on the miss probability, which is a function
of the miss distance.  Thus, calculating the costs in the Bayesian
decision criterion, $C_{ij}$, requires an assessment of the miss
distance of each particle at any time step during the engagement.
Previous works by the authors showed that DGL laws, specifically
DGL1~\cite{Mudrik2026unified} and the doubly compensated DGL law
(DGLCC)~\cite{Mudrik2026comprehensive}, are natural candidates for this
task, as the resulting miss distance of each law can be directly
inferred, at any time, from the game space of that law.  Thus, we
briefly present the construction of these game spaces and their usage
for the DGL laws.  \rev{We refer to the perfect-information
differential game underlying these laws as the DGL game.}

To construct the game space, we use the optimal controls to generate
the following dynamics for the optimal trajectories
\begin{equation}
	\dot{\bar{z}}(\tau) = \Gamma(\tau) \sign \bar{z}(0)
	\label{eq:z_dot_eq}
\end{equation}
where $\Gamma(\tau)$ is a known function of the game
parameters~\cite{glizer_complete_2008}
and $\bar{z}$ is the zero-effort miss (ZEM), which is the
miss distance that results if both players do not apply any further
acceleration commands until the end of the game. The ZEM is a scalar
nondimensional variable, which becomes an uncertainty set in the
delayed information case~\cite{Mudrik2026comprehensive}, and the
optimal controls are based on its center.

\begin{remark}
  Throughout this work we assume that the interceptor’s maneuver
  capability is superior to that of the evader, i.e., $\mu > 1$.  This
  assumption is reasonable, as it is well known that the interceptor’s
  options are very limited against an optimally evading, highly
  maneuverable target, unless the interceptor possesses a
  maneuverability advantage over it~\cite{shinar_solution_1981}.
\end{remark}

\begin{figure}[tbh]
  \centering
  \includegraphics[width=0.45\textwidth]{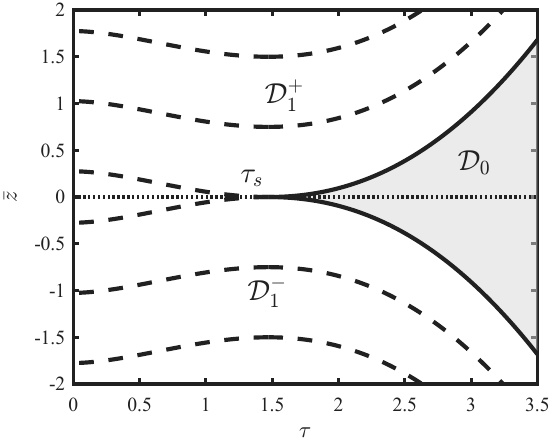}
\caption{Game space decomposition ($\mu > 1$).  Optimal trajectories,
  and boundaries of the singular region (solid lines).}
\label{fig:Game_Space_mu_1}
\end{figure}

Integrating \eqref{eq:z_dot_eq} yields the regular optimal
trajectories.  However, there are cases where these trajectories do
not fill the entire game space $(\bar{z},\tau)$; in such cases, the
game solution yields a decomposition of the game space into two
regions.  Figure~\ref{fig:Game_Space_mu_1} presents the shape of the
game space (for $\mu > 1$), where $\Gamma(\tau)$ has a unique positive
root $\tau_s$, which marks the apex of the singular region.

We decompose the regular region, denoted by $\mathcal{D}_{1}$, into
two subregions: 1) the upper regular region, for positive values of
the ZEM, denoted by $\mathcal{D}_{1}^{+}$, and 2) the lower regular
region, for negative values of the ZEM, denoted by
$\mathcal{D}_{1}^{-}$.  The optimal feedback strategies in the regular
region are
\begin{equation}
	\bar{u}^{*}(\tau) = \bar{v}^{*}(\tau) = \sign \bar{z}(\tau).
	\label{eq:opt_strat}
\end{equation}
In the singular region, $\mathcal{D}_{0}$, the optimal strategies of
both players are arbitrary.  For practical reasons we use a linear,
chattering prevention strategy for the pursuer inside the singular
region~\cite{shaferman_stochastic_2016}, yielding
\begin{equation}
	\bar{u}^{*}(\bar{z},\tau) = 
	\begin{cases}
		\sign \bar{z}(\tau),                                      & (\bar{z},\tau) \in \mathcal{D}_{1} \\
		\text{sat} \frac{\bar{z}(\tau)}{k \bar{z}^{*}(\tau)}, & (\bar{z},\tau) \in \mathcal{D}_{0}
	\end{cases}
	\label{eq:DGL_GL}
\end{equation}
where $\sat(\cdot)$ stands for the saturation function, $0 < k \leq 1$
is the portion of the singular region in which the control is linear,
and $\bar{z}^{*}$ is the boundary of the singular region.

The resulting miss distance of the perfect information differential
game, which is the unique value of the game in dimensional form, is
\begin{align}
	\label{eq:J}
  M_{s}&(\bar{z},\tau) = \notag \\ 
       &\begin{cases}
         a_{T}^{\max} \tau_{M}^{2}[\abs{\bar{z}(\tau)} + \int_{\tau}^{0} \Gamma(\tau')d\tau'], & (\bar{z},\tau) \in \mathcal{D}_{1} \\
         a_{T}^{\max} \tau_{M}^{2}[\int_{\tau_{s}}^{0} \Gamma(\tau')d\tau'],                                & (\bar{z},\tau) \in \mathcal{D}_{0}
	\end{cases}
\end{align}
Note that $M_s$ is identical for any initial condition within the
singular region, assuming both players employ their optimal
strategies therein.

\subsection{\rev{Invariance of the Game Solution}}
\label{subsec:invariance}

\rev{%
The methodology developed in the sequel replaces the terminal cost of
the DGL game of
Sec.~\fullref{sec:SSKP_awa_GL}{subsec:game_space}, the miss distance,
by the miss probability that the warhead's lethality model assigns to
it.  This replacement does not require solving the game anew.
Pure-strategy equilibria of a game depend only on the players'
ordering of the outcomes~\cite{Osborne2004}, and the miss probability
\eqref{eq:P_m}, being strictly increasing in the miss distance,
preserves that ordering.  The following proposition specializes this
known property to the DGL game.

\begin{proposition}
  \label{prop:invariance}
  Let $g : [0,\infty) \to \mathbb{R}$ be nondecreasing.  Then every
  saddle point of the perfect-information DGL game is a saddle point
  of the game having its terminal cost
  $M_{s}$ replaced by $g(M_{s})$, whose value is accordingly $g$
  applied to \eqref{eq:J}.  If, moreover, $g$ is strictly increasing,
  the two games have the same saddle-point strategies
  \eqref{eq:opt_strat}, and hence the same optimal trajectories
  \eqref{eq:z_dot_eq} and the same game space decomposition.  Because
  the miss probability \eqref{eq:P_m} is strictly increasing in the
  miss distance, the latter holds, in particular, for $g = P_{m}$.
\end{proposition}

\begin{proof}
  Let $(\bar{u}^{*},\bar{v}^{*})$ be a saddle point of the game having
  the terminal cost $M_{s}$, so that
  \begin{equation}
    \label{eq:saddle}
    M_{s}(\bar{u}^{*},\bar{v}) \leq
    M_{s}(\bar{u}^{*},\bar{v}^{*}) \leq M_{s}(\bar{u},\bar{v}^{*})
  \end{equation}
  for all admissible $\bar{u}$ and $\bar{v}$.  A nondecreasing $g$
  preserves both inequalities, rendering
  $(\bar{u}^{*},\bar{v}^{*})$ a saddle point of the game having the
  terminal cost $g(M_{s})$; the value of a zero-sum game being unique,
  it is $g(M_{s}(\bar{u}^{*},\bar{v}^{*}))$.  A strictly increasing $g$
  reflects the inequalities as well, so that a saddle point of the
  modified game is one of the original, and the two sets coincide.
  Integrating \eqref{eq:z_dot_eq} under the common strategies yields
  common optimal trajectories, and with them the common decomposition
  of the game space into the regions those trajectories do and do not
  fill.  The argument involves only the ordering of the outcomes, no
  regularity of $M_{s}$ being required, and is thus valid throughout
  the game space, the boundary of the singular region included.
\end{proof}

For a differentiable and strictly increasing $g$, this result can also be
seen from the Hamiltonian of the game~\cite{shinar_solution_1981},
which is affine in the players' controls.  The optimal strategies then
depend on the terminal cost only through the sign of its gradient, and
$\partial g(M_{s})/\partial \bar{z} = g'(M_{s}) \, \partial
M_{s}/\partial \bar{z}$ has the sign of $\partial M_{s}/\partial
\bar{z}$.  Moreover, the two costs have the same level sets, and the
boundary of the singular region is one of them.  Hence,
Fig.~\ref{fig:Game_Space_mu_1} describes both games, and the feedback
law \eqref{eq:DGL_GL} is unchanged.  The lethality model superimposes
on this decomposition only the locus of the warhead's effective lethal
radius, as shown in Fig.~\ref{fig:Examp_game_space}.

The second part of Proposition~\ref{prop:invariance} does not hold if
$g$ is only nondecreasing.  An example is the CC model
\eqref{eq:cookie_damage}, which is flat on either side of its
lethality radius.  Under it, every strategy that keeps the miss
distance below that radius is optimal.  The set of saddle points then
strictly contains that of the original game, and the game space
decomposition is not preserved.  The PLM \eqref{eq:P_m} is strictly
increasing, which is a further advantage over the CC model.

The main implication of Proposition~\ref{prop:invariance} is that
under perfect information, the lethality model does not affect the
guidance command.  In contradistinction, under imperfect information
it does, because the
Bayesian risk \eqref{Eq:I} averages the costs over the posterior PDF
and therefore depends on their values, which $g$ changes.%
}
\subsection{Maximizing the Kill Probability}
\label{subsec:Bayes4int}
In previous works, the authors applied Bayesian decision theory to
generate EA variants of the
DGL1~\cite{Mudrik2026unified} and DGLCC~\cite{Mudrik2026comprehensive}
laws, keeping the final miss distance as the cost function.  In
contradistinction, our approach in this work sets the miss probability
as the cost function, thereby directly maximizing the interceptor's
kill probability using its known warhead lethality model.  In the
following, we define the cost function and the decision criterion's
hypotheses.
\subsubsection{Cost Function}
\label{subsubsec:cost}
Because the Bayesian decision criterion minimizes risk, we set the
cost function to be the miss probability \eqref{eq:J_Pm}, evaluated
via the lethality model \eqref{eq:P_m}.  The cost $C_{ij}$ therefore
represents the miss probability that results when the guidance law
assumes, over a finite horizon, that hypothesis $H_{i}$ is true while
hypothesis $H_{j}$ is, in fact, true.  Each cost is computed from the
game space: in the regular region, the miss distance follows from the
first case of \eqref{eq:J}; in the singular region, it is constant,
and given by the second case of \eqref{eq:J}.  The miss distance is
then mapped to a miss probability via \eqref{eq:P_m}, where $\mu_w$
and $\sigma_w$ are the given warhead lethality parameters.

\begin{figure}[tbh]
\centering
  \includegraphics[width=0.45\textwidth]{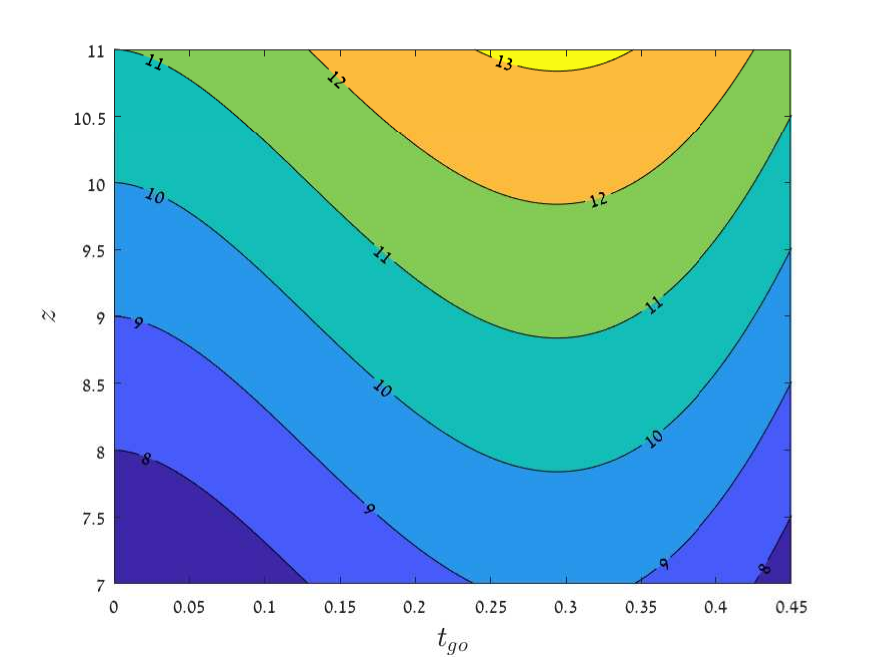}
  \caption{Miss distance cost function over a dimensional game space
    with marked equicost lines.}
  \label{fig:Game_Space_MD}
\end{figure}

\begin{figure}[tbh]
  \centering
  \includegraphics[width=0.45\textwidth]{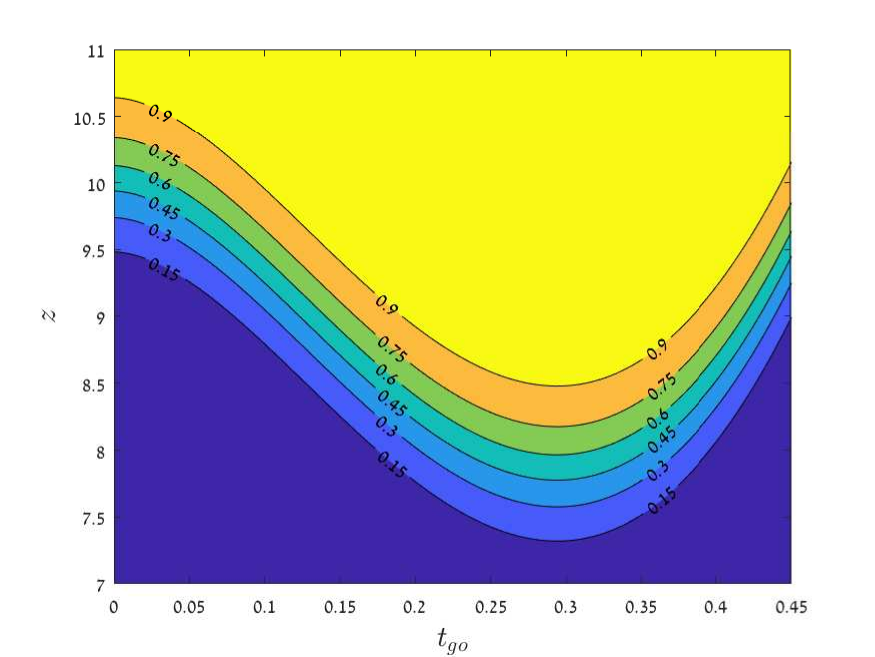}
  \caption{Miss probability cost function over a dimensional game
    space with marked equicost lines.}
  \label{fig:Game_Space_Pm}
\end{figure}

Figures~\ref{fig:Game_Space_MD} and \ref{fig:Game_Space_Pm} illustrate
the two cost functions over a dimensional game space.  When the miss
distance serves as the cost, Fig.~\ref{fig:Game_Space_MD} shows that
the cost increases linearly with the ZEM for any given time-to-go,
which clearly follows from the fact that, in this case, the lethality
model is not taken into
account. 

On the other hand, when the miss probability serves as the cost (with
lethality model parameters $\mu_w = 10$ and $\sigma_w = 1/2$),
Fig.~\ref{fig:Game_Space_Pm} shows that the penalty is highly
nonlinear: at the end of the engagement ($t_{go}=0$), increasing the
miss distance from seven to eight meters incurs negligible additional
cost, whereas increasing it from nine to ten meters has a dramatic
effect, as the miss probability jumps from nearly $0$ to approximately
$0.5$.
\rev{By Proposition~\ref{prop:invariance}, the equicost curves of
  Figs.~\ref{fig:Game_Space_MD} and~\ref{fig:Game_Space_Pm} are level
  sets of the same function, the miss distance.  The figures differ in
  the levels drawn and in their labels: with $\mu_w = 10$ and
  $\sigma_w = 1/2$, the curves of Fig.~\ref{fig:Game_Space_Pm}
  correspond to miss distances between $9.5$ and $10.6$~m, whereas
  those of Fig.~\ref{fig:Game_Space_MD} span $8$ to $13$~m.  The
  Bayesian decision criterion is affected by these cost values, which
  differ between the two cases.}
%
\subsubsection{Hypotheses}
\label{subsubsec:hypo}
Based on the guidance law's game space
(Fig.~\ref{fig:Game_Space_mu_1}) and the assumption that $u_{T}$ is
drawn from a known set of $R$ maneuvers, we define the following $R+2$
hypotheses:
\begin{description}
\item[$H_{1}$:]~The state is inside the upper regular region.
\item[$H_{k+1}$:]~The state is inside the singular region, and the
  target is in mode $k$, $k \in \{1,...,R\}$.
\item[$H_{R+2}$:]~The state is inside the lower regular region.
\end{description}  
\begin{remark}
  Explicitly addressing the case where the target performs bang-bang
  evasion maneuvers ($R=2$), \cite{Mudrik2026unified} considers only
  four hypotheses.
\end{remark}
\begin{remark}
  If the normalized time-to-go of a particle $j$ satisfies
  $\tau_{j} < \tau_{s}$, it belongs
  to $H_{1}$ if its ZEM is positive, and to $H_{R+2}$ if it is
  negative. If the ZEM is exactly zero (an event of measure zero since
  $\bar{z} \in \mathbb{R}$), the particle is assigned to $H_{1}$ for
  completeness.
\end{remark}

Applying the Bayesian decision criterion in~\eqref{Eq:Decision} to the
interception problem requires the following calculations using the
IMMPF's particle states and weights from the current and previous
time-steps.

\subsubsection{Likelihood Probabilities}
\label{subsubsec:likelihood}
Given hypothesis $H_j$, the likelihood of the measurements is
approximated by summing the IMMPF weights of the particles belonging
to that hypothesis~\cite{Mudrik2026unified}:
\begin{equation}
  \label{eq:likelihood}
  P(\meas \given H_{j})
  = \sum_{j'\in H_{j}} {w}^{j'}_{k}, \quad j \in \{1, \dots,m\}.
\end{equation}
\subsubsection{Prior Probabilities}
\label{subsubsec:prior_probs}
The prior probabilities are calculated from the measurement history
$\meas_{k-1}$ available at time $t_{k-1}$, before the new measurement
is acquired at $t_{k}$. Let SW denote the event that the evader has
switched its acceleration command in $[t_{k-1},t_{k}]$, and let NSW
denote the complementary event. Then, the law of total probability
yields
\begin{align}
	P_{j}  &= \Pr(H_{j} \mid \meas_{k-1},\text{SW}) \Pr(\text{SW}) \notag \\
	&+ \Pr(H_{j} \mid \meas_{k-1},\text{NSW}) \Pr(\text{NSW}).
	\label{eq:prior}
\end{align}
The conditional hypothesis probabilities are obtained by propagating
forward the particle cloud from $t_{k-1}$ according to whether a
switch has occurred or not. The probabilities $\Pr(\text{SW})$ and
$\Pr(\text{NSW})$ are computed via the law of total probability,
conditioning on the evader's mode at $t_{k-1}$ and the transition
probability matrix (TPM), as detailed in~\cite{Mudrik2026unified}.
\subsubsection{Costs}
\label{subsubsec:costs}
The costs are computed from the miss probability \eqref{eq:P_m}, which
depends on the miss distance of each particle via \eqref{eq:J}.  We
distinguish between four cases:
	\begin{enumerate}
	\item The cost of correctly choosing hypothesis $H_{j}$ when $P(\meas \given H_{j})>0$ is 
	\begin{equation}
		C_{jj} =  \sum_{j'\in H_{j}} \tilde{w}^{j'}_{k}P_{m}
		\left( M_{s} \left( \bar{z}_{j'},\tau_{j'} \right) \right)
		\label{eq:c_jj}
	\end{equation}
	where $\bar{z}_{j'}$ and $\tau_{j'}$ are the normalized ZEM and time-to-go of particle $j' \in H_{j}$, respectively,
	and $\tilde{w}^{j'}_{k}$ is the corresponding normalized weight at time $t_{k}$:
	\begin{equation}
		\tilde{w}^{j'}_{k} = \frac{w^{j'}_{k}}{\sum_{j'\in H_{j}} {w}^{j'}_{k}}.
	\end{equation}
	
	Unlike the miss-distance-based cost
        of~\cite{Mudrik2026unified,Mudrik2026comprehensive}, which is
        identically zero inside the singular region, the miss
        probability $P_m(M_s)$ is strictly positive there (since
        $P_m(M_s) > 0$ for any finite miss distance under the PLM),
        rendering the lethality model
        consequential to the Bayesian decision process even when the
        game state lies inside the singular region.
      \item The cost of wrongly choosing hypothesis $H_{i}$ when
        $P(\meas \given H_{i}) >0$, when hypothesis $H_{j}$ is true
        and $P(\meas \given H_{j})>0$, is
	\begin{equation}
		C_{ij} =  \sum_{j'\in H_{j}}
		\tilde{w}^{j'}_{k} \sum_{i'\in H_{i}} \tilde{w}^{i'}_{k}   P_{m}
		\left( M_{s} \left( \bar{z}_{i'j'},\tau_{j'} - h \right) \right)
		\label{eq:c_ij1}
	\end{equation}
	where $\bar{z}_{i'j'}$ is the normalized ZEM of particle
        $j' \in H_{j}$, propagated over the interval
        $[\tau_{j'},\tau_{j'}-h]$ using the acceleration command
        from~\eqref{eq:DGL_GL}.  The command corresponds to particle
        $i' \in H_{i}$ and is held constant over the prediction horizon
        $h$ (in normalized time).
	
      \item The cost of wrongly choosing hypothesis $H_{i}$ when
        $P(\meas \given H_{i})=0$, when hypothesis $H_{j}$ is true and
        $P(\meas \given H_{j})>0$, is
	\begin{equation}
          C_{ij}= \sum_{j'\in H_{j}} \tilde{w}^{j'}_{k}P_{m}
          \left( M_{s} \left( \bar{z}_{i'j'},\tau_{j'} - h \right) \right)
		\label{eq:c_ij2}
	\end{equation}
	
      \item If $P(\meas \given H_{j}) = 0$, then $C_{jj}$ and $C_{ij}$
        can be set arbitrarily, per the optimal Bayesian decision rule
        of~\eqref{Eq:I}.
\end{enumerate}
\subsection{KPM Guidance Law}
\label{subsec:gst_gl}
The Bayesian decision criterion~\eqref{Eq:Decision} yields an optimal
decision when the unnormalized additional risks $I_i$, computed over a
finite horizon $h_{\max}$, are nonzero. When all $I_i$ are zero, any
acceleration command can be used without significantly affecting
interception performance; in this case, we default to the
deterministic guidance law of~\eqref{eq:DGL_GL}.
\begin{remark}
  When the optimal Bayesian acceleration command is nonunique,
  trajectory shaping can improve other aspects of the interceptor's
  operation, such as the information it
  acquires\cite{shaviv_estimation-guided_2017,Mudrik2026unified}.
\end{remark}

When the $I_i$ are nonzero, the resulting KPM variant of a
differential game-based guidance law becomes
\begin{equation}
	\label{eq:ModGL}
	\bar{u}=
	\begin{cases}
		+ 1          & H_1 \text { is decided} \\
		\sum_{i'\in H_{2}}\tilde{w}_{k}^{i'}\frac{z_{i'}( t_{go;i'} )}{z_{i'}^{*}( t_{go;i'})} & H_{2} \text { is decided}\\ 
		\vdots & \\
		
		\sum_{i'\in H_{R+1}}\tilde{w}_{k}^{i'}\frac{z_{i'}( t_{go;i'} )}{z_{i'}^{*}( t_{go;i'})} & H_{R+1} \text { is decided}\\ 
		- 1         & H_{R+2} \text { is decided} 
	\end{cases}.
\end{equation}
The KPM law has $R+2$ modes, corresponding to the $R+2$ hypotheses.
The first mode applies maximal acceleration toward the upper regular
region, while the last mode applies it toward the lower regular
region.  The remaining $R$ modes correspond to singular-region
hypotheses, where the acceleration command is the weighted mean of the
chattering prevention commands from~\eqref{eq:DGL_GL} of the particles
belonging to the chosen hypothesis.
\subsection{Illustrative Example}
\label{subsec:illustrative}
\begin{figure}[tbh]
  \centering
  \includegraphics[width=0.45\textwidth]{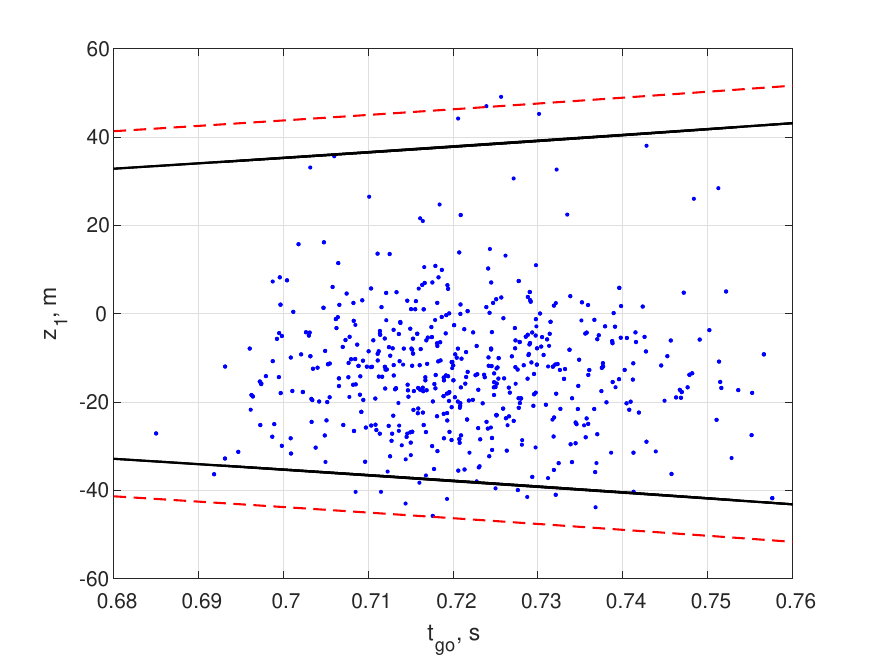}
  \caption{Posterior PDF\rev{,} represented by 1000 equally weighted
    particles (blue dots)\rev{, superimposed} over the dimensional game
    space of DGL1.
    Singular region boundaries: solid black lines; boundaries of
    warhead high kill probability region: dashed red lines.}
  \label{fig:Examp_game_space}
\end{figure}

To further highlight the differences between the proposed KPM law and
the common miss-distance-minimizing solution,
Fig.~\ref{fig:Examp_game_space} shows the posterior PDF, discretely
represented by 1000 equally-weighted particles spread over the
dimensional DGL1 game space.  The scenario parameters are
$a_{M}^{\max}=45$~g, $a_{T}^{\max}=20$~g, and
$\tau_{M}=\tau_{T}=0.2$~s, so that the interceptor can guarantee HTK
performance in the deterministic case.  The target performs a
bang-bang evasion maneuver with a single switch ($R=2$).  The Bayesian
decision criterion is applied with both cost functions for a
single-time-step horizon of $0.01$~s, using a warhead with $\mu_w=10$
and $\sigma_w = 1/2$.  Both cost functions are evaluated using the
same warhead parameters; only the guidance cost function differs.  The
dashed red lines in the figure demarcate the region where
$M_s \leq \reff(3) = 8.5~\text{m}$, corresponding to a kill
probability of $0.9987$.

Table~\ref{table:1} presents the unnormalized additional risks $I_i$
for the four hypotheses under both cost functions.  Under the miss
distance cost, $H_{3}$ is optimal because more particles (16) lie in
the lower regular region than in the upper one (4).  Under the
miss-probability cost, however, $H_1$ is optimal: two of the four
upper-region particles lie outside the high-kill-probability zone (and
the other two are near its boundary), whereas no lower-region particle
crosses its boundary and only one resides near it.  This example
highlights a key distinction between the two approaches---under the
miss distance cost, every particle outside the singular region
contributes equally to the risk, whereas under the miss-probability
cost, only those particles near or beyond the warhead's effective
lethal radius $\reff(3)$ incur significant cost.
\begin{table}[thb]
  \caption{Bayesian decision criterion parameters.}
  \centering
  \begin{tabular}{l c c c c}
    \hline\hline
    Cost          & $I_{1}$ & $I_{2}$ & $I_{3}$ & $I_{4}$  \\ \hline 
    $|M_{s}|$, m  & $1.99$  & $1.14$  & $2.1\cdot 10^{-4}$ & $5.8\cdot 10^{-2}$      \\
    $P_{m}$       & $8.4\cdot 10^{-6}$  & $9\cdot 10^{-6}$  & $1.1\cdot 10^{-5}$  & $1.7\cdot 10^{-5}$      \\
    \hline\hline
  \end{tabular}	
  \label{table:1}
\end{table}

\section{Simulation Study}
\label{sec:sim}
The performance of the proposed guidance strategy is evaluated via a
numerical simulation study.  The interception scenario
follows~\cite{shaferman_stochastic_2016}, and we also use the MC
results of~\cite{Mudrik2026comprehensive}.  We first present the
simulation scenario and design considerations, then compare the
regular, EA, and KPM variants of the DGL1 and DGLCC laws via MC
simulations. The EA variants follow \cite{Mudrik2026unified} and
\cite{Mudrik2026comprehensive} for the DGL1 and DGLCC, respectively.
\subsection{Simulation Scenario}
\label{subsec:param}
A ballistic missile defense scenario is considered, comprising a
single interceptor and a single highly maneuverable target that
performs a bang-bang evasion maneuver with a single command switch
($R=2$).
\rev{The optimality of the bang-bang evasion maneuver is supported
  by~\cite{Mudrik2026bangbang}, where the set of optimal evasion
  strategies in realistic, stochastic settings is shown to include a
  bang-bang policy.}
The target is initialized in the $-Y_{I}$ direction with
$\gamma_{T}(0)=-\pi/2$~rad, and the interceptor's velocity vector
points toward the initial target location. The maneuver capabilities
are $a_{M}^{\max}=45$~g and $a_{T}^{\max}=20$~g, with time constants
$\tau_{M}=\tau_{T}=0.2$~s and speeds $V_{M}=V_{T}=2500$~m/s, yielding
a nominal engagement time of $3$~s. The measurement noise is
$\nu \sim \mathcal{N}(0,\sigma^{2})$ with $\sigma=0.5$~mrad, and the
infrared sensor sampling rate is $f=100$~Hz.

\subsection{Design Considerations}
Each guidance law (DGL1 and DGLCC) is tested in three variants:
regular (deterministic), EA (miss-distance-minimizing), and KPM
(kill-probability-maximizing).  Following~\cite{Mudrik2026unified} and
\cite{Mudrik2026comprehensive}, the DGL1 and DGLCC laws require
warheads with high kill probabilities up to miss distances of $15$~m
and $10$~m, respectively.  We therefore evaluate performance using the
following four lethality models: 1) an HTK warhead ($\mu_w=1/2$,
$\sigma_w=0.01$), approximating a step function at half a meter; 2) a
`small' warhead ($\mu_w=5$, $\sigma_w=1/2$), which would probably
prove insufficient for both laws; 3) a `medium' warhead ($\mu_w=10$,
$\sigma_w=1/2$), which should be sufficient for the DGLCC law but not
for the DGL1; and 4) a `large' warhead ($\mu_w=15$, $\sigma_w=1/2$),
which should be sufficient for both laws. The prediction horizon is
$h_{\max} = 0.01$~s (one sensor time step).

The IMMPF uses $4000$ weighted particles ($2000$ per mode), initialized with equal mode probabilities. The initial state estimate satisfies
\begin{equation}
	\hat{x}_{R}\sim\mathcal{N}(\bar{x}_{R},P_{R})
	\label{eq:RadarInitPDF}
\end{equation}
where $\bar{x}_{R}$ is the true initial state of~\eqref{eq:StateVec1}, and the initial covariance matrix is
\begin{equation}
	P_{R}=\text{diag} \{ 50^{2},(1\pi/180)^{2},(3\pi/180)^{2},10^{2} \}.
	\label{eq:P_R}
\end{equation}
The DGLCC law additionally uses a fixed-lag smoother with online time-delay estimation, as detailed in~\cite{Mudrik2026comprehensive}.

The TPM is
\begin{equation}
	\Pi = \begin{bmatrix}
		0.999 & 0.001 \\
		0.001 & 0.999
	\end{bmatrix}.
\end{equation}
The small transition probabilities reduce estimation bias during periods without acceleration switches, at the cost of slower detection of evasion maneuver changes. The resulting increase in estimation time-delays is compensated by the proposed strategy.

\subsection{Monte Carlo Analysis}
The six guidance law variants are compared via MC simulations against
two types of targets.  The nominal target performs a bang-bang evasion
maneuver with a switch time uniformly distributed over $[0,3]$~s.
Following~\cite{Mudrik2026comprehensive}, the interval $[1.5,2.5]$~s
is the most challenging for the interceptor; we therefore define a
`smart' target to be one that switches within this interval.

Table~\ref{tab:SSKP} presents the empirical SSKP evaluated based on
$3000$ MC runs against the nominal target.  DGL1-KPM achieves the
highest SSKP across all warhead types, substantially improving over
both the EA and regular variants.  All three DGLCC variants perform
poorly for the HTK and small warheads: DGLCC lacks HTK
capability by construction, and for the small warhead the EA and KPM
variants are outperformed by the regular variant because they do not
account for nonrational target behavior, as reported
in~\cite{Mudrik2026comprehensive}.  When the warhead is sufficient,
however, DGLCC-KPM achieves the highest SSKP among the DGLCC variants.

\begin{table}[tbh]
  \caption{Empirical SSKP of guidance law/lethality model
    combinations.  Nominal target.}
	\label{tab:SSKP}
\centering
\begin{tabular}[\linewidth]{ll|cccc}
\hline
\multicolumn{2}{l}{Guidance system} &       \multicolumn{4}{c}{Warhead type}         \tabularnewline \hline
Law   & Variant                     & HTK & Small   & Medium      & Large    \tabularnewline \hline\hline
      & Regular                     & $0.611$     & $0.725$ & $0.812$     & $0.902$  \tabularnewline
DGL1  & EA                          & $0.644$     & $0.764$ & $0.85$      & $0.938$  \tabularnewline
      & KPM                         & $0.674$     & $0.835$ & $0.959$     & $0.999$  \tabularnewline \hline
      & Regular                     & $0.011$     & $0.547$ & $0.927$     & $0.978$  \tabularnewline
DGLCC & EA                          & $0.01 $     & $0.391$ & $0.936$     & $0.997$  \tabularnewline
      & KPM                         & $0.013$     & $0.382$ & $0.957$     & $0.998$  \tabularnewline \hline\hline
\end{tabular}
\end{table}

Table~\ref{tab:SSKP2} presents the empirical SSKP evaluated based on
$1000$ MC runs against the smart target.  As could be expected, SSKP
performance degrades across all laws, but the ranking shifts:
DGLCC-KPM now achieves the best SSKP for medium and large warheads, as
it degrades less than DGL1-KPM under challenging evasion timing.
DGL1-KPM remains superior for the HTK and small warheads,
where all DGLCC variants perform poorly---in particular, the target
evades with probability one against every DGLCC variant with the
HTK warhead.  Thus, DGLCC-KPM is preferred when the warhead is
at least medium-sized, while DGL1-KPM is preferred when warhead
sufficiency cannot be guaranteed.

Among DGL1 variants, the KPM variant degrades the least, retaining an
SSKP of $0.886$ with a medium-sized warhead and nearly unity with a
large warhead.  Similarly, DGLCC-KPM with medium-sized and large
warheads is the least affected by the smart target's timing, achieving
the highest overall SSKP with minimal degradation relative to the
nominal scenario.
\begin{table}[tbh]
  \caption{Empirical SSKP of guidance law/lethality model
    combinations.  `Smart' target.}
\label{tab:SSKP2}
\centering
\begin{tabular}[\linewidth]{ll|cccc}
\hline
\multicolumn{2}{l}{Guidance system} &                      \multicolumn{4}{c}{Warhead type}                         \tabularnewline \hline
Law   & Variant                     & HTK & Small               & Medium             & Large                \tabularnewline \hline\hline
      & Regular                     & $0.34$     & $0.437$  & $0.527$  & $0.728$    \tabularnewline
DGL1  & EA                          & $0.437$    & $0.536$  & $0.646$  & $0.831$   \tabularnewline
      & KPM                         & $0.468$    & $0.635$  & $0.886$  & $0.994$    \tabularnewline \hline
      & Regular                     & $0$        & $0.38$   & $0.862$  & $0.957$    \tabularnewline
DGLCC & EA                          & $0$        & $0.35$   & $0.903$  & $0.994$    \tabularnewline
      & KPM                         & $0$        & $0.365$  & $0.931$  & $0.997$    \tabularnewline \hline\hline
\end{tabular}
\end{table}

\subsection{\rev{Sensitivity to Warhead Parameter Uncertainty}}
\label{subsec:robust}

\rev{%
We examine the sensitivity to errors in the lethality model.  The interceptor's warhead is known,
but a realistic lethality model also depends on the vulnerability of
the target, on the terminal geometry of the interception, and on the
detonation point set by the proximity
fuze~\cite{shinar_what_2003,ShinarForteKantor1994}.  Against a
nonnominal target, the parameters $\mu_w$ and $\sigma_w$ are, therefore,
known only approximately.  The error of practical concern is the
optimistic one, in which the true lethal radius is smaller than the
assumed one.  We therefore retain the three DGL1 variants, since
DGL1-KPM is the law preferred when warhead sufficiency cannot be
guaranteed.  The DGL1-KPM law is designed for the medium warhead, and
the three variants are evaluated against a true warhead whose lethal
radius is drawn below the assumed value and whose transition width is
drawn about it.

Table~\ref{tab:robust} presents the resulting SSKP against the smart
target.  All three variants degrade, since the true warhead is less
capable than the assumed one.  DGL1-KPM degrades the most, by $3.0$
percentage points against some $2.4$ for the EA and regular variants,
as could be expected: it alone is matched to the lethality model, and
so, of the three variants, it is the most sensitive to a model
mismatch.  However, that additional half point is negligible compared
with the advantage of DGL1-KPM, which retains a lead of $35$
percentage points over the regular variant and $23$ over the EA
variant.  For the case examined, these results demonstrate robustness
of the proposed paradigm with respect to inexact knowledge of the
warhead's model.%
}

\begin{table}[tbh]
  \caption{\rev{Empirical SSKP of the DGL1 variants under uncertainty
      in the warhead parameters.  `Smart' target; the KPM variant is
      designed for the medium warhead.}}
\label{tab:robust}
\centering
\rev{%
\begin{tabular}[\linewidth]{l|cc}
\hline
Variant & Nominal warhead  & Uncertain warhead                         \tabularnewline \hline
        & $\mu_w=10$       & $\mu_w\sim\mathcal{U}(8,10)$              \tabularnewline
        & $\sigma_w=1/2$   & $\sigma_w\sim\mathcal{U}(1/4,1)$          \tabularnewline \hline\hline
Regular & $0.527$          & $0.502$                                   \tabularnewline
EA      & $0.646$          & $0.622$                                   \tabularnewline
KPM     & $0.886$          & $0.856$                                   \tabularnewline \hline\hline
\end{tabular}%
}
\end{table}

\section{Conclusions}
\label{sec:concl}
This work presents a guidance strategy that replaces the traditional
miss-distance cost with the warhead's miss probability, thereby
directly maximizing the interceptor's \rev{SSKP}.  Complying
with the GST, the Bayesian decision-theoretic framework exploits the
game space decomposition to make optimal decisions over the posterior
PDF of the game state.  \rev{Replacing the miss-distance cost by the
miss probability is shown to leave the saddle-point strategies of the
underlying perfect-information game unaltered, so that the lethality
model affects the guidance command only through the posterior PDF of
the game state.}  The resulting KPM variants of DGL1 and DGLCC
are validated via extensive Monte Carlo simulations against nominal
and nonnominal (`smart') targets with four warhead types.  Because the
nonlinear miss-probability cost assigns disproportionate weight to
particles near the warhead's effective lethal radius, rather than
penalizing all miss distances equally, the KPM variants
improve the SSKP over both the regular and EA variants, \rev{DGL1-KPM
for every warhead type considered and DGLCC-KPM for the medium and
large warheads.  A sensitivity study indicates that this advantage is
maintained when the warhead's lethality model is known only
approximately}.  The DGLCC law lacks HTK capability by construction, a
limitation that the KPM modification cannot overcome.  Notably,
because the proposed methodology modifies only the guidance law, it
can be deployed as a software update to existing interceptors without
altering the warhead or other hardware, thereby extending their
operational effectiveness against evolving threats.  Against a smart
target, DGL1-KPM is preferred when the warhead lethality is
insufficient, while DGLCC-KPM is preferred when the warhead is at
least medium-sized, as it degrades less under challenging evasion
timing.

\bibliographystyle{IEEEtaes}
\bibliography{Bib1,references}
\end{document}